\documentclass[prd,aps,reprint,noeprint,superscriptaddress,amsmath, amssymb]{revtex4-2}
\usepackage{graphicx}
\usepackage{subfigure}
\usepackage{epsfig}
\usepackage{multirow}
\usepackage{overpic}
\usepackage{color}
\usepackage{bm} 
\usepackage{xspace} 
\usepackage{dcolumn}
\usepackage{booktabs,tabularx}
\usepackage{array}
\usepackage{textcomp}
\usepackage[colorlinks,linkcolor=blue,urlcolor=blue,citecolor=blue]{hyperref}
\usepackage{tikz}
\usepackage{changes}
\usepackage{soul}

\usepackage{cleveref}

\newcommand{\ee}{e^+e^-}
\newcommand{\Lambdabar}{\bar{\Lambda}}
\newcommand{\Omegabar}{\bar{\Omega}^+}

\sethlcolor{green}
\setstcolor{red}

\begin{document}
\normalsize
\parskip=5pt plus 1pt minus 1pt


\title{\boldmath Measurements of the absolute branching fractions of $\Omega^-$ decays and test of the $\Delta I = 1/2$ rule}

\author{
{\small M.~Ablikim$^{1}$, M.~N.~Achasov$^{5,b}$, P.~Adlarson$^{74}$, X.~C.~Ai$^{80}$, R.~Aliberti$^{35}$, A.~Amoroso$^{73A,73C}$, M.~R.~An$^{39}$, Q.~An$^{70,57}$, Y.~Bai$^{56}$, O.~Bakina$^{36}$
% ...
\\
\vspace{0.2cm}
(BESIII Collaboration)\\
\vspace{0.2cm} {\it
$^{1}$ Institute of High Energy Physics, Beijing 100049, People's Republic of China\\
% ...
$^{5}$ Budker Institute of Nuclear Physics SB RAS (BINP), Novosibirsk 630090, Russia\\
% ...
$^{35}$ Johannes Gutenberg University of Mainz, Johann-Joachim-Becher-Weg 45, D-55099 Mainz, Germany\\
$^{36}$ Joint Institute for Nuclear Research, 141980 Dubna, Moscow region, Russia\\
% ...
$^{39}$ Liaoning Normal University, Dalian 116029, People's Republic of China\\
% ...
$^{56}$ Southeast University, Nanjing 211100, People's Republic of China\\
$^{57}$ State Key Laboratory of Particle Detection and Electronics, Beijing 100049, Hefei 230026, People's Republic of China\\
% ...
$^{70}$ University of Science and Technology of China, Hefei 230026, People's Republic of China\\
% ...
$^{73}$ University of Turin and INFN, (A)University of Turin, I-10125, Turin, Italy; (B)University of Eastern Piedmont, I-15121, Alessandria, Italy; (C)INFN, I-10125, Turin, Italy\\
$^{74}$ Uppsala University, Box 516, SE-75120 Uppsala, Sweden\\
% ...
$^{80}$ Zhengzhou University, Zhengzhou 450001, People's Republic of China\\
\vspace{0.2cm}
% ...
$^{b}$ Also at the Novosibirsk State University, Novosibirsk, 630090, Russia\\
% ...
}\vspace{0.4cm}  }
} 
\begin{abstract}
    Based on a data set of $(27.12\pm0.10)\times 10^8$ $\psi(3686)$ events collected
    at the BESIII experiment, the absolute branching fractions of the
    three dominant $\Omega^-$ decays are measured to be
    $\mathcal{B}_{\Omega^- \to \Xi^0 \pi^-} =
    (25.03\pm0.44\pm0.53)\%$, $\mathcal{B}_{\Omega^- \to \Xi^- \pi^0}
    = (8.43\pm0.52\pm0.28)\%$, and $\mathcal{B}_{\Omega^- \to \Lambda
      K^-} = (66.3\pm0.8\pm2.0)\%$, where the first and second
    uncertainties are statistical and systematic, respectively.  The
    ratio between $\mathcal{B}_{\Omega^- \to \Xi^0 \pi^-}$ and
    $\mathcal{B}_{\Omega^- \to \Xi^- \pi^0}$ is determined to be
    $2.97\pm0.19\pm0.11$, which is in good agreement with the PDG
    value of $2.74\pm0.15$, but greater by more than four standard
    deviations than the theoretical prediction of 2 obtained from
    the $\Delta I = 1/2$ rule.  
\end{abstract}

\maketitle


Although the $\Omega^-$ baryon was discovered 60 years
ago~\citep{Barnes:1964pd}, its spin was not experimentally determined
until 2006~\cite{BaBar:2006omx}, under the assumption that the spin of
$\Xi^0_c$ is  1/2.
The first model-independent determination of the $\Omega^-$ spin was conducted by 
the BESIII collaboration in 2021~\cite{BESIII:2020lkm}, 
and many of its properties still remain unknown.  
The $\Omega^-$ baryon plays a unique role in the baryon
family as the only decuplet baryon that solely decays weakly.  The
weak decays of the octet baryons and the decays of $\Omega^-$ provide
important information about the interplay between the strong and weak
interactions~\cite{Goswami:1970wqc, Carone:1991ni, Tandean:1998ch,
  Egolf:1998vj, Antipin:2007ya}.

Although isospin is not conserved in weak interactions, there is an
experimentally well-established $\Delta I = 1/2$ rule, which states
that in weak interactions, the $\Delta I = 1/2$ amplitude is strongly
enhanced, while the $\Delta I = 3/2$ amplitude is suppressed.  In
general, this rule is well satisfied~\cite{Georgi:1984zwz,
  Jenkins:1991bt, Cirigliano:2011ny, Salone:2022lpt}.  For example,
the $\Lambda \to p \pi^-$ and $\Lambda \to n \pi^0$ branching
fractions (BFs)~\cite{ParticleDataGroup:2022pth} imply that the
$\Delta I = 3/2$ amplitude in $\Lambda$ decays is less than
$2\%$~\cite{Overseth:1969bxc}.  The only significant violation of the
$\Delta I = 1/2$ rule is observed in $\Omega^-$ decays.  The BFs of the
three dominant decays of $\Omega^-$ listed by the Particle Data Group
(PDG)~\cite{ParticleDataGroup:2022pth} were measured by the
CERN-WA-046 experiment nearly 40 years ago~\cite{BGHORS:1984jku}.
Using the 40-year-old data, the ratio between the BFs of $\Omega^- \to
\Xi^0 \pi^-$ and $\Omega^- \to \Xi^- \pi^0$ ($\mathcal{B}_{\Omega^-
  \to \Xi^0 \pi^-}/\mathcal{B}_{\Omega^- \to \Xi^- \pi^0}$) is
$2.74\pm0.15$~\cite{ParticleDataGroup:2022pth}, while the value
predicted by the $\Delta I =1/2$ rule is 2~\cite{Carone:1991ni,
  Mommers:2022dgw}.  These results have never been confirmed by any
other experiment, leading to some skepticism~\cite{Jenkins:1991bt,
  Tandean:1998ch, Salone:2022lpt}.  There are two points of view: one
suggests abandoning the $\Delta I = 1/2$ rule, and obtaining more data
to make phenomenologically reliable predictions~\cite{Tandean:1998ch};
the other assumes the $\Delta I = 1/2$ rule to be true and that new
measurements will overturn the previous results~\cite{Mommers:2022dgw,
  Li:2016tlt}.  To resolve this, new measurements of the BFs for
  $\Omega^-$ decays are urgently needed.  With the world's largest
  $\psi(3686)$ data sample, BESIII has an excellent opportunity to
  measure the absolute BFs of $\Omega^-$ decays and test the $\Delta
  I = 1/2$ rule.


In this Letter, we utilize a double-tag (DT) method to measure the
absolute BFs of the $\Omega^- \to \Xi^0 \pi^-$, $\Omega^- \to \Xi^-
\pi^0$, and $\Omega^- \to \Lambda K^-$ decays (unless otherwise noted,
the charge-conjugated decays are always implied).  The single-tag (ST)
events of $\bar{\Omega}^+$ baryons are reconstructed via the decay
$\bar{\Omega}^+ \to \bar{\Lambda} K^+$.  The events where a signal
candidate can be reconstructed from the particles recoiling against
the ST $\bar{\Omega}^+$ baryon are called DT events.  To improve the
detection efficiencies, only the $\pi^-$, $\pi^0$, and $K^-$ mesons
from $\Omega^-$ decays are reconstructed on the signal side for the
three $\Omega^-$ decay channels.  The BF of a signal
decay is determined by
\begin{equation}
    \label{eq:BFcalcu}
    \mathcal{B}_{\rm sig} = {N}_{\rm DT}~\epsilon_{\rm ST}/({N}_{\rm ST}~\epsilon_{\rm DT}) = {N}_{\rm DT}/({N}_{\rm ST}~\epsilon_{\rm sig}),
\end{equation}
where ${N}_{\rm ST}$ and ${N}_{\rm DT}$ are the ST and DT
yields, respectively. The $\epsilon_{\rm sig} = \epsilon_{\rm
  DT}/\epsilon_{\rm ST}$ is the signal efficiency in the presence of
an ST $\bar{\Omega}^+$ baryon, where $\epsilon_{\rm ST}$ and
$\epsilon_{\rm DT}$ are the ST and DT efficiencies, respectively.

The analysis is based on a sample of $(27.12 \pm 0.10) \times 10^8$
$\psi(3686)$ events~\cite{BESIII:2017tvm} collected with the BESIII
detector at the BEPCII collider.  The BESIII
detector~\cite{Ablikim2010} records symmetric $e^+e^-$ collisions
provided by the BEPCII storage ring~\cite{Yu:2016cof}, which operates
with a peak luminosity of $1\times10^{33}$~cm$^{-2}$s$^{-1}$ in the
center-of-mass energy range from 2.0 to 4.95~GeV.  BESIII has
collected large data samples in this energy
region~\cite{BESIII:2020nme, Huang:2022wuo}.  The cylindrical core of
the BESIII detector covers 93\% of the full solid angle and consists
of a helium-based multilayer drift chamber~(MDC), a plastic
scintillator time-of-flight system~(TOF), and a CsI(Tl)
electromagnetic calorimeter~(EMC), which are all enclosed in a
superconducting solenoidal magnet providing a 1.0~T magnetic
field. The solenoid is supported by an octagonal flux-return yoke with
resistive plate counter muon identification modules interleaved with
steel.  The charged-particle momentum resolution at $1~{\rm GeV}/c$ is
$0.5\%$, and the $dE/dx$ resolution is $6\%$ for electrons from Bhabha
scattering. The EMC measures photon energies with a resolution of
$2.5\%$ ($5\%$) at $1$~GeV in the barrel (end cap) region. The time
resolution in the TOF barrel region is 68~ps, while that in the end
cap region was 110~ps. The end cap TOF system was upgraded in 2015 using multigap
resistive plate chamber technology, providing a time resolution of 60 ps, which benefits about 83\%
of the data used in this analysis~\cite{Li:TOF, Guo:TOF, Cao:2020ibk}.

Monte Carlo (MC) simulation samples are used to optimize the event
selection and estimate the background.  The simulation is performed by
the {\sc geant4}-based~\cite{G42002iii} BESIII software
system~\cite{etde_20820412}, which includes the geometric description
of the BESIII detector and the detector response.  The simulation
models the beam energy spread and initial state radiation in the
$e^+e^-$ annihilations with the generator {\sc
  kkmc}~\cite{Jadach2001}.  The known decay modes of $\psi(3686)$ are
modeled with {\sc evtgen}~\cite{Lange2001,*Ping2008}, and the
remaining unknown decays are modeled with {\sc
  lundcharm}~\cite{Chen2000,*Yang2014a}.  Final state radiation from
charged final state particles is incorporated using {\sc
  photos}~\cite{Richter-Was:1992hxq}.  The signal MC samples,
$\Omega^- \to$ $X$, used to determine the ST efficiency, and $\Omega^-
\to \Xi^0(\to X) \pi^-$, $\Omega^- \to \Xi^-(\to X) \pi^0$, and
$\Omega^- \to \Lambda(\to X) K^-$, used to determine the DT efficiency,
are generated uniformly in phase space. Final state $X$ indicates
inclusive decay, and each sample consists of 2.54 million events.

We first measure the absolute BFs for the decay channels $\Omega^- \to \Xi^0 \pi^-$ and
$\Omega^- \to \Xi^- \pi^0$.  Charged tracks detected in the
MDC are required to be within a polar angle ($\theta$) range of
$|\!\cos\theta|<0.93$, where $\theta$ is the polar angle with respect to the
symmetry axis of the MDC.  Particle
identification (PID) for charged track combines measurements of the
d$E/$d$x$ in the MDC and the flight time in the TOF to form a
likelihood $\mathcal{L}(h)(h = p, K, \pi)$ for each hadron $h$
hypothesis.  Charged tracks with $\mathcal{L}(p)>\mathcal{L}(K)$,
$\mathcal{L}(p)>\mathcal{L}(\pi)$ and $\mathcal{L}(p)>0.001$ are
identified as protons, and those with
$\mathcal{L}(K)>\mathcal{L}(\pi)$ are identified as kaons.  The
remaining charged tracks are assigned as pions by default.

For $\Lambdabar$ candidates,
the $\bar{p} \pi^+$ pairs are constrained to have a common vertex, and
the invariant mass of a $\bar{p} \pi^+$
combination is required to be within $[1.111, 1.121]$~GeV/$c^2$.
Vertex fits are performed to the $\Lambdabar K^+$ pairs to improve the
mass resolution of the $\Omegabar$ candidates.  If there is more than
one $\Omegabar$ candidate, the one with the minimum value of $\Delta E
= |E_{\Omegabar} - E_{\rm beam}|$ is kept for further analysis, where
$E_{\Omegabar}$ is the energy of the reconstructed $\Omegabar$
candidate in the $\ee$ center-of-mass system and $E_{\rm beam}$ is the
beam energy.  In addition, the invariant mass of $\Lambdabar K^+$
($M_{\Lambdabar K^+}$) must lie in the $\Omegabar$ signal region of
$[1.664, 1.680]$~GeV/$c^2$.

\begin{figure}[htbp]
    \begin{center}
        \mbox{
            \put(-125, 0){
                \begin{overpic}[width = 1.0\linewidth]{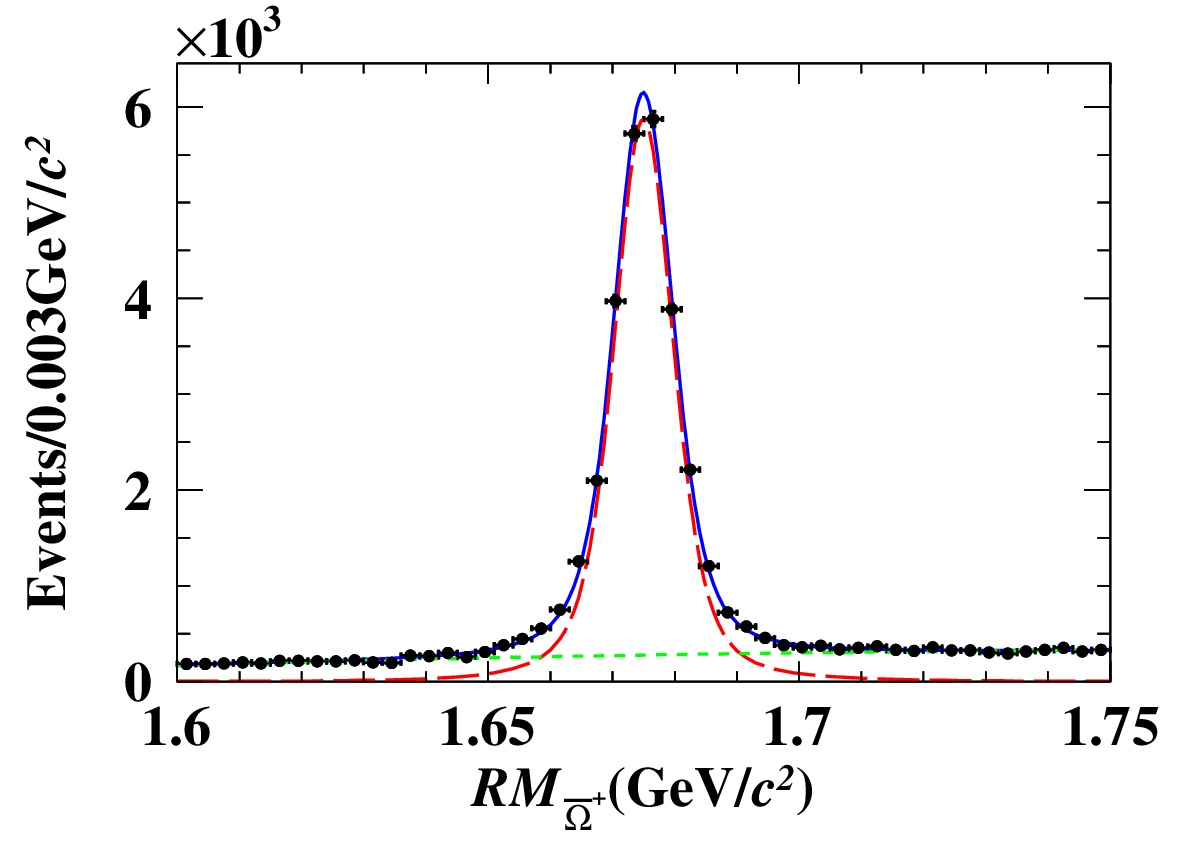}
                \end{overpic}
            }
        }
    \end{center}
    \caption{Fit to the $RM_{\Omegabar}$ distribution. The dots with error bars are data, the blue solid line is the total fit, 
    and the green short-dashed and magenta long-dashed lines represent the fitted background and signal shapes, respectively.
    }
    \label{fig:SingleTag}
\end{figure}

Events in the $\Omegabar$ sideband region, defined as $M_{\Lambdabar
  K^+} \in [1.648, 1.656]$ $\cup$ $[1.688, 1.696]$~GeV/$c^2$, are used
to study the potential peaking backgrounds in data. No peaking
background is found.  To determine the ST yields, an unbinned
maximum-likelihood fit is performed to the recoil-mass spectrum
against the reconstructed $\Omegabar$ ($RM_{\Omegabar}$), as shown in
Fig.~\ref{fig:SingleTag}.  In the fit, the signal shape is described
by the MC simulated shape convolved with a Gaussian
function with free parameters, where the Gaussian function is used to
compensate for the difference in mass resolution between data and MC
simulation.  The background shape is described by a second-order
Chebyshev polynomial.  The signal region of $RM_{\Omegabar}$ is
defined as [1.652, 1.695]~GeV/$c^2$, and the number of ST $\Omegabar$
baryons is $25819 \pm 188$, with a corresponding efficiency of
$23.55\%$, as listed in Table~\ref{tab:BFs}.

\begin{table}[htbp]
    \caption{The ST yields ($\rm{N}_{\rm ST}$), ST efficiency
    ($\epsilon_{\rm ST}$), DT yields ($\rm{N}_{\rm DT}$), DT
    efficiency ($\epsilon_{\rm DT}$) and the absolute BFs of the three
    dominant
    $\Omega^-$ decays, $\Omega^- \to \Xi^0 \pi^-$, $\Omega^- \to \Xi^-
    \pi^0$, and $\Omega^- \to \Lambda K^-(\Omegabar \to \bar{\Lambda}
    K^+)$.
    Here the uncertainties are statistical only.}
    \label{tab:BFs}
    \setlength{\extrarowheight}{1.0ex}
     \renewcommand{\arraystretch}{1.0}
    \begin{center}
        \scalebox{0.82}{
        \begin{tabular} {l | c | c | c | c | c}
            \hline \hline
            Decay mode                                     & $\rm{N}_{\rm ST}$ & $\epsilon_{\rm ST} (\%)$ & $\rm{N}_{\rm DT}$ & $\epsilon_{\rm DT} (\%)$ & BF (\%)    \\
            \hline
            $\Omega^- \to \Xi^0 \pi^-$ & \multirow{2}{1.8cm}{\centering $25819\pm188$} & $\multirow{2}{1.1cm}{\centering 23.55}$ & $5411\pm95$ & $19.72$ & $25.03\pm 0.44$\\
            \cline{1-1} \cline{4-6}
            $\Omega^- \to \Xi^- \pi^0$ & &  & $794 \pm 49$ & $8.59$ & $8.43\pm0.52$\\
            \cline{1-6}
            $\Omega^- \to \Lambda K^-$ & $12111 \pm 127$ & $22.78$ & $4877 \pm 72$ & $13.64$ & $67.25\pm0.99$\\
            $\bar{\Omega}^+ \to \bar{\Lambda} K^+$ & $13705 \pm 139$ & $24.31$ & $5427 \pm 78$ & $14.75$ & $65.26\pm0.94$\\
            \hline \hline
        \end{tabular}}
    \end{center}
\end{table}

The $\pi^{-}(\pi^0)$ candidates from the decay $\Omega^- \to \Xi^{0}
\pi^{-}(\Xi^{-} \pi^0)$ are reconstructed in the recoil side of the
$\Omegabar$.  The tracks that satisfy $|\cos
\theta| < 0.93$ and $\mathcal{L}(\pi)>\mathcal{L}(K)$ are regarded as
pion candidates.  The photons used to reconstruct the $\pi^0$
candidates are detected in the EMC.
Each photon is required to have an EMC energy deposit of more than
25~MeV in the barrel region ($|\cos\theta| < 0.80$) or more than
50~MeV in the end-cap region ($0.86<|\cos\theta|<0.92$).  To suppress
electronic noise and showers unrelated to the event, the difference
between the EMC time and the event start time is required to be within
(0, 700)~ns.  Furthermore, to reject showers that originate from
charged tracks, the angle between the shower and its closest charged
track must be greater than $10^{\circ}$.  The photons are combined
into pairs and at least one photon is required to come from the barrel
region.  The photon pairs with invariant mass within the range
$[0.115, 0.150]$ GeV/$c^2$ are denoted as $\pi^0$ candidates.  A
kinematic fit~\cite{Yan:2010zze} constraining the $\gamma \gamma$
invariant mass to the known $\pi^0$
mass~\cite{ParticleDataGroup:2022pth} is performed, and the resulting
$\chi^2$ must be less than 200.  If there is more than one
$\pi^-(\pi^0)$ candidate, the one with the highest energy is kept for
further analysis.

For the measurement of $\mathcal{B}_{\Omega^- \to \Xi^- \pi^0}$, it is
necessary to veto the background from the decays $\Omega^- \to \Lambda
K^-$ and $\Omega^- \to \Xi^0 \pi^-$, which also have $\pi^0$s in the
signal side.  To veto the background from $\Omega^- \to \Xi^0 \pi^-$,
if the highest momentum track on the signal side carries the opposite
charge to the tagged $\Omegabar$, it is assigned as a pion and the
recoil-mass against the $\Omegabar \pi^-$ system, $RM_{\Omegabar
  \pi^-}$, is calculated.  We require $RM_{\Omegabar \pi^-} >
1.38$~GeV/$c^2$ to exclude events from $\Omega^- \to \Xi^0 \pi^-$,
where there would be a peak around the known $\Xi^0$ mass in the
$RM_{\Omegabar \pi^-}$ spectrum.  For the background from $\Omega^-
\to \Lambda K^-$, if there are $K^-$ tracks on the signal side, the
$K^-$ with the highest $\mathcal{L}(K)$ is used to calculate the
recoil-mass against the $\Omegabar K^-$ system, $RM_{\Omegabar K^-}$,
and $RM_{\Omegabar K^-} > 1.2$~GeV/$c^2$ is required to veto this
background.

Potential peaking backgrounds in the measurements of
$\mathcal{B}_{\Omega^- \to \Xi^0 \pi^-}$ and $\mathcal{B}_{\Omega^-
  \to \Xi^- \pi^0}$ are investigated by analyzing the inclusive MC
sample and the events in the $M_{\Lambdabar K^+}$ or $RM_{\Omegabar}$
sideband regions from data. The $RM_{\Omegabar}$ sideband region is
defined as $RM_{\Omegabar} \in [1.608, 1.630]~\cup~[1.718,
  1.739]$~GeV/$c^2$.  The number of peaking background events is
estimated to be $46 \pm 20$ for the $\Omega^- \to \Xi^0 \pi^-$ mode,
while there is no peaking background for the $\Omega^- \to \Xi^-
\pi^0$ mode.

\begin{figure}[htbp]
    \begin{center}
        \mbox{
            \put(-130, 0){
                \begin{overpic}[width = 1.0\linewidth]{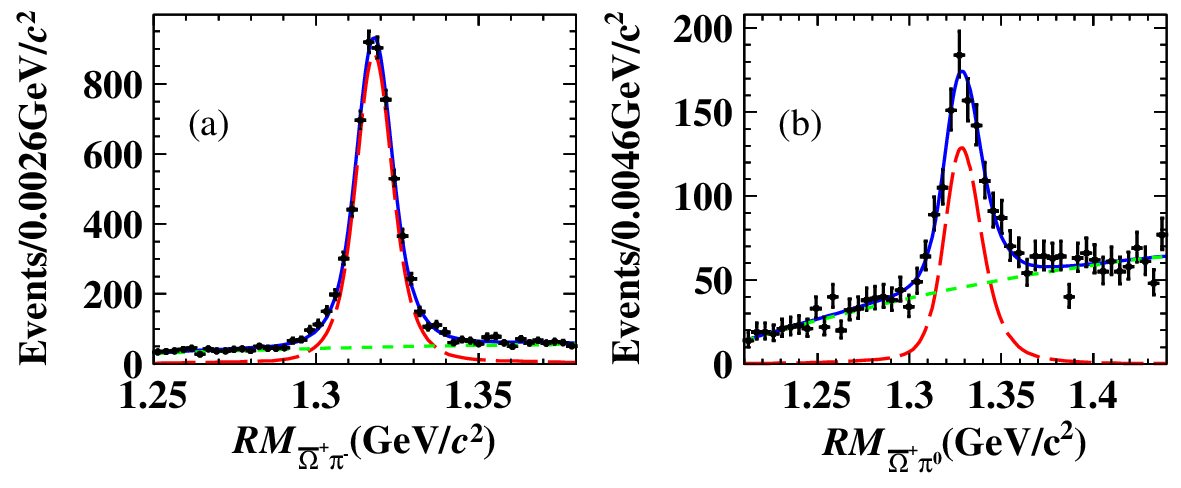}
                \end{overpic}
            }
        }
    \end{center}
    \caption{Fits to the distributions of (a) $RM_{\Omegabar \pi^-}$
      and (b) $RM_{\Omegabar \pi^0}$. The dots with error bars are
      data, the blue solid line is the total fit, and the green
      short-dashed and magenta long-dashed lines represent the fitted
      background and signal shapes, respectively.  }
    \label{fig:XipiDT}
\end{figure}

For the signal events of the DT sample, 
the recoil-mass spectrum against the $\Omegabar \pi^- (\Omegabar \pi^0)$ system, 
$RM_{\Omegabar \pi^-}(RM_{\Omegabar \pi^0})$, peaks around the $\Xi^0(\Xi^-)$ mass.
An unbinned maximum-likelihood fit is performed on 
$RM_{\Omegabar \pi^-}(RM_{\Omegabar \pi^0})$ to determine the DT yields, as shown in
Fig.~\ref{fig:XipiDT}.  In the fit, the signal shape is described by
the MC simulated shape convolved with a Gaussian function.  To obtain
the MC simulated shape of $RM_{\Omegabar \pi^0}$, a truth matching
method is used, where the opening angle between the reconstructed and
MC-truth momentum directions of $\pi^0$ is required to be less than
$10^{\circ}$ and the energy difference between them is required to be
less than 0.1~GeV.  The background shape is described by a
second-order Chebyshev polynomial.  After subtracting the number of
peaking background events, the number of DT events of the $\Omegabar
\pi^- (\Omegabar \pi^0)$ sample is determined to be $5411 \pm 95~(794
\pm 49)$, as listed in Table~\ref{tab:BFs}.  The polar angle
and energy distributions of $\pi^0$ in data from $\Omega^- \to \Xi^-
\pi^0$ are well described by the PHSP model.  The polar angle
distribution of $\pi^-$ from $\Omega^- \to \Xi^0 \pi^-$ is also
consistent between data and MC simulation, while the transverse
momentum distributions of $\pi^-$ ($p_{T, \pi^-}$) is not.  To obtain
a more accurate DT efficiency in the measurement of
$\mathcal{B}_{\Omega^- \to \Xi^0 \pi^-}$, the events of the signal MC
sample are weighted according to the $p_{T,
  \pi^-}$ distribution in data. The weight factors are the ratios $n^i_{\rm
  data}/n^i_{\rm MC}$ which are obtained in different $p_{T,
  \pi^-}$ bins, where $n^i_{\rm data}$ and $n^i_{\rm MC}$ are the
numbers of DT $\Omegabar \pi^-$ candidates in the $i$-th $p_{T,
  \pi^-}$ bin from data and MC samples, respectively.  The resultant
DT efficiencies are  $19.72 \%$ and
$8.59\%$ for $\Omegabar \pi^-$ and $\Omegabar \pi^0$, respectively,
as listed in Table~\ref{tab:BFs}.

The absolute BFs of $\Omega^- \to \Xi^0 \pi^-$ and $\Omega^- \to \Xi^- \pi^0$
are determined to be $\mathcal{B}_{\Omega^- \to \Xi^0 \pi^-} = (25.03
\pm 0.44)\%$ and $\mathcal{B}_{\Omega^- \to \Xi^- \pi^0} = (8.43 \pm
0.52)\%$, respectively, as listed in Table~\ref{tab:BFs}.
Here, the uncertainties are statistical only.

For the $\Omega^- \to \Lambda K^-$ channel, the BFs
($\mathcal{B}_{\Omega^- \to \Lambda K^-}$) of the charge conjugate
modes $\Omega^- \to \Lambda K^-$ and $\Omegabar \to \Lambdabar K^+$
are determined separately, since we observe the same decay mode of
$\Omega \to \Lambda K$ on both the tag and signal sides.  In this channel,
the selection criteria of the ST events are the same as those in the
$\Omega^- \to \Xi^0 \pi^- (\Xi^- \pi^0)$ BF measurement.  The ST yield
is determined to be $12111 \pm 127~(13705 \pm 139)$, and the ST
efficiency is $22.78\%~(24.31\%)$ for the tagged
$\Omegabar(\Omega^-)$, as listed in Table~\ref{tab:BFs}.

For the signal side, at least one good charged $K^- (K^+)$ track is
required.  If there is more than one $K^- (K^+)$, the one with the
highest $\mathcal{L}(K)$ is kept for further study.  Potential peaking
backgrounds for the DT events in the measurement of
$\mathcal{B}_{\Omega^- \to \Lambda K^-}$ are investigated by analyzing
the inclusive MC sample and the events in the $M_{\Lambdabar
  K^+}(M_{\Lambda K^-})$ or $RM_{\Omegabar}(RM_{\Omega^-})$ sideband
regions from data.  We find $22 \pm 7 (24 \pm 9)$ peaking background
events for the DT $\Omegabar K^-(\Omega^- K^+)$ sample.

\begin{figure}[htbp]
    \begin{center}
        \mbox{
            \put(-130, 0){
                \begin{overpic}[width = 1.0\linewidth]{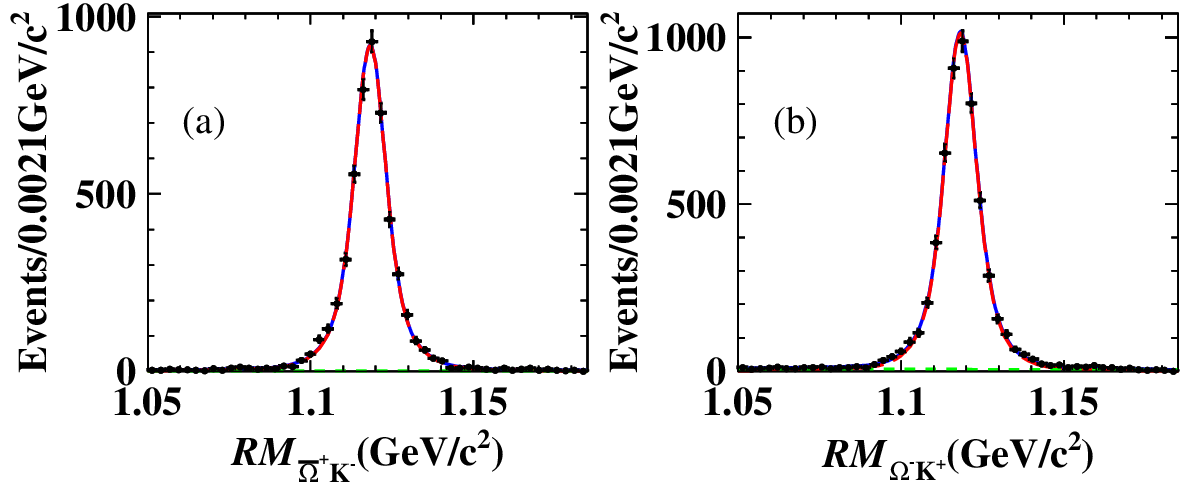}
                \end{overpic}
            }
        }
    \end{center}
    \caption{Fits to the distributions of (a) $RM_{\Omegabar K^-}$ and (b) $RM_{\Omega^- K^+}$. The dots with error bars are data, the blue solid line is the total fit, 
    and the green short-dashed and magenta long-dashed lines represent the fitted background and signal shapes, respectively.
    }
    \label{fig:LamKDT}
\end{figure}

For the signal events of the DT sample, 
the recoil-mass spectrum against the $\Omegabar K^-(\Omega^- K^+)$ system, 
$RM_{\Omegabar K^-}(RM_{\Omega^- K^+})$, should peak around the $\Lambda(\bar{\Lambda})$ mass.
Therefore, an unbinned maximum-likelihood fit is performed on 
$RM_{\Omegabar K^-}(RM_{\Omega^- K^+})$ to determine the DT yield, as shown in
Fig.~\ref{fig:LamKDT}. In the fit, the signal shape is described by
the MC simulated shape convolved with a Gaussian
function.  The background shape is described by a first-order
Chebyshev polynomial.  After subtracting the number of peaking
background events, the number of DT events of $\Omega^- \to \Lambda
K^- (\Omegabar \to \Lambdabar K^+)$ is determined to be $4877 \pm
72~(5427 \pm 78)$, as listed in Table~\ref{tab:BFs}.  To obtain a more
accurate DT efficiency, the events of the PHSP MC sample are weighted
according to the observed distribution of $K^-(K^+)$ transverse
momentum.  The resulting DT efficiency is determined to be
$13.64\%~(14.75\%)$, as listed in Table~\ref{tab:BFs}.

The absolute BFs of the charge separated decays $\Omega^- \to \Lambda K^-$ and
$\Omegabar \to \Lambdabar K^+$ are determined to be
$\mathcal{B}^{\Omega^-}_{\Lambda K^-} = (67.25 \pm 0.99) \%$ and
$\mathcal{B}^{\Omegabar}_{\Lambdabar K^+} = (65.26 \pm 0.94)\%$,
respectively, as listed in Table~\ref{tab:BFs}.  Weighting
these two BFs and considering the correlation of the 1292 overlap
events between the $\bar \Omega^+K^-$ and $\Omega^-K^+$
samples~\cite{James:2006zz}, we obtain the average BF of
$\mathcal{B}_{\Omega^- \to \Lambda K^-} = (66.3 \pm 0.8)\%$, where the
uncertainty is statistical only.

With the DT method, most uncertainties related to the ST selection
cancel.  The sources of the systematic uncertainties are summarized in
Table~\ref{tab:systematic}. Each of them is described in the following
paragraphs.
\begin{table}[htbp]
    \caption{BF Relative systematic uncertainties in \%.}
    \label{tab:systematic}
    \setlength{\extrarowheight}{1.0ex}
     \renewcommand{\arraystretch}{1.0}
    \begin{center}
    \scalebox{0.95}{
        \begin{tabular} {l c c c}
            \hline \hline
            Source & $\mathcal{B}_{\Omega^- \to \Xi^0 \pi^-}$ & $\mathcal{B}_{\Omega^- \to \Xi^- \pi^0}$ & $\mathcal{B}_{\Omega^- \to \Lambda K^-}$\\
            \hline
            Photon detection                            & -     & $2.0$ & -    \\
            $\pi^0$ reconstruction                      & -     & $1.0$ & -    \\
            Pion/kaon tracking                          & $0.3$ & -     & $1.0$           \\
            Pion/kaon PID                               & $0.2$ & -     & $1.0$           \\
            ST signal shape                             & $0.9$ & $0.9$ & $0.9$                \\
            DT signal shape                             & $0.0$ & $1.4$ & $0.2$                \\
            ST background shape                         & $0.7$ & $0.7$ & $0.7$           \\
            DT background shape                         & $0.2$ & $0.9$ & $0.8$                \\
            ST background fluctuation                   & $0.4$ & $0.4$ & $0.4$     \\
            Weighting procedure                         & $1.7$ & -     & $2.2$           \\
            $\Xi^0\pi^-$ veto                           & -     & $1.3$ &-  \\
            Truth match                                 & -     & $0.5$ &-  \\
            MC statistics                               & $0.1$ & $0.2$ & $0.1$  \\
            \hline \hline
            Total                                       & $2.1$ & $3.3$ & $3.0$                \\
            \hline \hline
        \end{tabular}
    }
    \end{center}
\end{table}

The systematic uncertainty associated with the photon reconstruction
efficiency is estimated to be 1.0\% per photon~\cite{BESIII:2010mhh},
while that due
to the $\pi^0$ reconstruction is also 1.0\%~\cite{BESIII:2010ank}.  The
uncertainties arising from the tracking and PID efficiencies are both
1.0\% per kaon track~\cite{BESIII:2018ldc}.  A control sample $J/\psi
\to p \bar{p} \pi^+ \pi^-$ is used to estimate the uncertainties of
the pion tracking and PID efficiencies.  The efficiency differences
between data and MC simulation for the control sample are used to
re-weight the signal MC sample of $\Omega^- \to \Xi^0 \pi^-$.  The
differences between the nominal and re-weighted detection efficiencies
are taken as the systematic uncertainties, which are 0.3\% and 0.2\%
for pion tracking and PID efficiencies, respectively.

The uncertainties associated with the ST and DT signal shapes are
estimated by replacing the Gaussian resolution function with a
double-Gaussian function.  The differences in the signal yields are
taken as the systematic uncertainties, which are 0.9\% for the ST
yield, and 0.0\%, 1.4\%, and 0.2\% for the DT yields of $\Omega^- \to
\Xi^0 \pi^-$, $\Omega^- \to \Xi^- \pi^0$, and $\Omega^- \to \Lambda
K^-$, respectively.  The uncertainty due to the ST background shape is
studied by changing the second-order Chebyshev polynomial to a
first-order or third-order Chebyshev polynomial, and the largest
difference on the ST yield, 0.7\%, is taken as the uncertainty.  The
uncertainties caused by the DT background shapes are investigated by
increasing the order of the nominal Chebyshev polynomial by one, and
the changes of the DT yields are taken as the systematic
uncertainties, which are 0.2\%, 0.9\%, and 0.8\% for $\Omega^- \to
\Xi^0 \pi^-$, $\Omega^- \to \Xi^- \pi^0$, and $\Omega^- \to \Lambda
K^-$, respectively.  In addition, the uncertainty due to the
background fluctuation of the ST yield, 0.4\%, is also considered as a
systematic uncertainty.

To obtain reliable signal efficiencies, the signal MC samples of
$\Omega^- \to \Xi^0 \pi^-$ and $\Omega^- \to \Lambda K^-$ are weighted
to match the data.  To estimate the associated systematic uncertainty,
the weight factors are randomly changed within one standard deviation
in each bin one thousand times to re-obtain the DT efficiencies.  The
distributions of the resulting DT efficiencies are fit with Gaussian
functions, and their standard deviations are taken as the systematic
uncertainties, which are 1.7\% and 2.2\% for $\Omega^- \to \Xi^0
\pi^-$ and $\Omega^- \to \Lambda K^-$, respectively.

The systematic uncertainty of vetoing the background of $\Omega^- \to
\Xi^0 \pi^-$ in the measurement of $\mathcal{B}_{\Omega^- \to \Xi^-
  \pi^0}$ is studied by varying this requirement from $RM_{\Omegabar
  \pi^-}>1.38$~GeV/$c^2$ to $RM_{\Omegabar \pi^-}>1.36$~GeV/$c^2$ or
$RM_{\Omegabar \pi^-}>1.40$~GeV/$c^2$.  The resulting largest
difference to the original BF, 1.3\%, is taken as the systematic
uncertainty.  Since there is no signal efficiency loss of vetoing the
background of $\Omega^- \to \Lambda K^-$, the systematic uncertainty
due to this requirement is negligible.

The systematic uncertainty of the truth matching method in the $\Omega^-
\to \Xi^- \pi^0$ mode originates from the requirements on the opening
angle between the reconstructed and MC-truth momentum directions of the
$\pi^0$, and the energy difference between the reconstructed and
MC-truth $\pi^0$s. The relevant uncertainties are estimated by varying
the opening angle value to be $5^{\circ}$ or $15^{\circ}$, and the
energy difference to be 0.05~GeV or 0.15~GeV. The resulting largest
difference to the original BF is assigned as the uncertainty, which is
0.5\%.

The uncertainties due to the MC statistics are estimated to be 0.1\%,
0.2\%, and 0.1\% for $\Omega^- \to \Xi^0 \pi^-$, $\Omega^- \to \Xi^-
\pi^0$, and $\Omega^- \to \Lambda K^-$, respectively.

Adding these systematic uncertainties in quadrature, we obtain the
total systematic uncertainties for the measurements of
$\mathcal{B}_{\Omega^- \to \Xi^0 \pi^-}$, $\mathcal{B}_{\Omega^- \to
  \Xi^- \pi^0}$, and $\mathcal{B}_{\Omega^- \to \Lambda K^-}$ to be
2.1\%, 3.3\%, and 3.0\%, respectively.

In summary, utilizing $(27.12\pm0.10)\times 10^8$ $\psi(3686)$ events
collected with the BESIII detector, the absolute BFs of $\Omega^- \to
\Xi^0 \pi^-$, $\Omega^- \to \Xi^- \pi^0$, and $\Omega^- \to \Lambda
K^-$ have been measured. The results are shown in
Table~\ref{tab:Results comparison}.  The ratio between $\mathcal{B}_{\Omega^-
\to \Xi^0 \pi^-}$ and $\mathcal{B}_{\Omega^- \to \Xi^- \pi^0}$ is
determined to be $2.97 \pm 0.19 \pm 0.11$, where the systematic
uncertainties associated with the ST yields cancel in the
calculation. Our result is consistent with the PDG value $2.74 \pm
0.15$ (combining statistical and systematic uncertainties in
quadrature), but differs from the expectation (equal to 2)
based on the $\Delta I = 1/2$ rule by more than four standard
deviations.  Our measurement and the PDG value both suggest a
surprisingly strong admixture of a $\Delta I = 3/2$ amplitude, which
is very different from the behavior observed in other measured weak decays,
especially the decays of the octet baryons~\cite{Tandean:1998ch}.  To
understand this phenomenon, the current effective-field-theory picture
about the interplay of the strong and weak interactions needs to be
improved~\cite{Mommers:2022dgw}.

\begin{table}[htbp]
    \caption{The obtained BFs (in \%) and comparison with the PDG values. The first and second uncertainties presented in this work are statistical and systematic, respectively.}
    \label{tab:Results comparison}
    \begin{center}
    \setlength{\extrarowheight}{1.0ex}
     \renewcommand{\arraystretch}{1.0}
    \scalebox{0.9}{
    \setlength{\extrarowheight}{1.0ex}
    \renewcommand{\arraystretch}{1.0}
    \vspace{0.2cm}
        \begin{tabular} {l c c c }
            \hline \hline
            BFs  & $\mathcal{B}_{\Omega^- \to \Xi^0 \pi^-}$ & $\mathcal{B}_{\Omega^- \to \Xi^- \pi^0}$ & $\mathcal{B}_{\Omega^- \to \Lambda K^-}$ \\
            \hline 
            This work& $25.03 \pm 0.44 \pm 0.53$ & $8.43 \pm 0.52 \pm 0.28$ & $66.3 \pm 0.8 \pm 2.0$\\ 
            PDG& $23.6 \pm 0.7 $ & $8.6 \pm 0.4$ & $67.8 \pm 0.7$\\
            \hline \hline
        \end{tabular}
    \vspace{-0.2cm}
    }
    \end{center}
\end{table}

\textbf{Acknowledgement}

The BESIII Collaboration thanks the staff of BEPCII and the IHEP computing center for their strong support. This work is supported in part by National Key R\&D Program of China under Contracts Nos. 2020YFA0406300, 2020YFA0406400; National Natural Science Foundation of China (NSFC) under Contracts Nos. 11635010, 11735014, 11835012, 11935015, 11935016, 11935018, 11961141012, 12022510, 12025502, 12035009, 12035013, 12061131003, 12192260, 12192261, 12192262, 12192263, 12192264, 12192265, 12221005, 12225509, 12235017; the Chinese Academy of Sciences (CAS) Large-Scale Scientific Facility Program; the CAS Center for Excellence in Particle Physics (CCEPP); Joint Large-Scale Scientific Facility Funds of the NSFC and CAS under Contract No. U1832207; CAS Key Research Program of Frontier Sciences under Contracts Nos. QYZDJ-SSW-SLH003, QYZDJ-SSW-SLH040; 100 Talents Program of CAS; The Institute of Nuclear and Particle Physics (INPAC) and Shanghai Key Laboratory for Particle Physics and Cosmology; ERC under Contract No. 758462; European Union's Horizon 2020 research and innovation programme under Marie Sklodowska-Curie grant agreement under Contract No. 894790; German Research Foundation DFG under Contracts Nos. 455635585, Collaborative Research Center CRC 1044, FOR5327, GRK 2149; Istituto Nazionale di Fisica Nucleare, Italy; Ministry of Development of Turkey under Contract No. DPT2006K-120470; National Research Foundation of Korea under Contract No. NRF-2022R1A2C1092335; National Science and Technology fund of Mongolia; National Science Research and Innovation Fund (NSRF) via the Program Management Unit for Human Resources \& Institutional Development, Research and Innovation of Thailand under Contract No. B16F640076; Polish National Science Centre under Contract No. 2019/35/O/ST2/02907; The Swedish Research Council; U. S. Department of Energy under Contract No. DE-FG02-05ER41374.

%

\end{document}